\documentclass{PoS}

\title{Pion masses in 2-flavor QCD with $\eta$ condensation}

\ShortTitle{Pion masses in 2-flavor QCD}

\author{\speaker{Sinya AOKI}\\
        Yukawa Institute for Theoretical Physics, Kyoto University, Kitashirakawa Oiwakecho, Sakyo-ku, Kyoto 606-8502, Japan\\
        E-mail: \email{saoki@yukawa.kyoto-u.ac.jp}}

\author{Michael CREUTZ\\
        Physics Department 510A, Brookhaven National Laboratory, Upton, NY 11973, USA\\
        E-mail: \email{mike@latticeguy.net}}

\abstract{We investigate the 2-flavor QCD with non-degenerate quark masses at
low-energy, using the chiral perturbation theory including the $\eta$ meson and anomaly effects.  
For the fixed $m_d\not=0$, the neutral pion becomes massless at two
values of $m_u$, between which a spontaneously CP broken phase appears with 
the neutral pion condensation.
We then show that the topological susceptibility  diverges at these two critical points. 
We also consider the case of
$m_u=m_d$ but $\theta=\pi$, equivalently  $m_u= - m_d$
with $\theta = 0$ by the chiral rotation and show that 
the CP symmetry is spontaneously broken by the $\eta$ condensation at small $m$. 
Around $m=0$, three pions become Nambu-Goldstone modes,  showing non-standard behavior that $m_\pi^2 = O(m^2)$, which, however, is consistent with the chiral Ward-Takahashi identities.}

\FullConference{The 32nd International Symposium on Lattice Field Theory,\\
		23-28 June, 2014\\
		Columbia University New York, NY}

\begin{document}

\section{Introduction}
"Massless up quark" seems a simple solution to  the strong CP problem, but  such a possibility is claimed to be already ruled out by the analysis based on lattice QCD simulations\cite{Nelson:2003tb}.
One of the present authors, however, has argued that a concept of massless up quark is ambiguous if all other quarks are massive\cite{Creutz:1995wf, Creutz:2003xu,Creutz:2003xc,Creutz:2005gb,Creutz:2013xfa},
since no chiral symmetry which guarantees masslessness of up quark exists in this case due to the chiral anomaly. Moreover there appears so-called a Dashen phase\cite{Dashen:1971aa}, where the CP symmetry is spontaneously broken, and the neutral pion becomes massless  at the phase boundaries, so that
the topological susceptibility diverges there.

In this report, we investigate above properties in 2-flavor QCD with non-degenerate quark masses, employing the chiral perturbation theory (ChPT) including the $\eta$ meson with anomaly effects\cite{Aoki:2014moa}.
Our ChPT analysis explicitly confirms  the above-mentioned properties.
We also analyze the case that $m_u=m_d=m$ but $\theta =\pi$, equivalently $m_u=-m_d$ with $\theta=0$, where
$m_u$ and $m_d$ are up and down quark masses, respectively. 
We show, deep in the CP violating phase with the $\eta$ condensation, that
three pions becomes Nambu-Goldstone (NG) bosons with the non-standard behavior as $m_\pi^2=O(m^2)$, which is, however, consistent with the chiral Ward-Takahashi identities (WTI).

\section{Phase structure and pion masses}
\subsection{ChPT with $\eta$ and anomaly}
We introduce the anomaly into the ChPT  at leading order (LO) as
\begin{eqnarray}
L =\frac{f^2}{2} {\rm tr}\, \left(\partial_\mu U \partial^\mu U^\dagger\right) -\frac{1}{2}{\rm tr}\, \left(M^\dagger U + U^\dagger M\right) -\frac{\Delta}{2}\left(\det\, U +\det\, U^\dagger\right), 
\label{eq:Lag}
\end{eqnarray}
where $f$ is the pion decay constant, $M$ is the quark mass matrix,  a field $U\in U(N_f)$ contains the flavor singlet $\eta$ meson as well as the on-singlet pions, and
a positive constant $\Delta$ gives an additional mass to the $\eta$.
For simplicity, $\det\, U$ terms with derivatives are neglected here since they do not change our main conclusions.

Assuming that quarks belong to the fundamental representation of SU($N_c$) color gauge group, the large $N_c$ argument\cite{Witten:1980sp,Rosenzweig:1979ay,Kawarabayashi:1980dp,Arnowitt:1980ne} leads to  the $\displaystyle\frac{\Delta}{N_c} (\log \det\, U)^2$ term instead of the third term in the above.
Rigorously speaking, however,
it s impossible to  uniquely determine the form of anomaly term in ChPT from the large $N_c$ argument,
since the fundamental and the 2-index anti-symmetric representations are identical at  $N_c=3$ but give very different anomaly terms in the large $N_c$ limit.
Therefore we use the above form for simplicity but also check our results with $(\log \det\, U)^2$.
 
\subsection{Warm-up: $N_f=1$ case}
As a warm-up example, let us consider the $N_f=1$ case. 
Since there is only one pseudo-scalar meson ($\eta$) which stays massive due to the anomaly,  
one may naively guess  
\begin{equation}
m_{\rm PS}^2 = \frac{\vert m \vert}{f^2}  + \delta m^2
\end{equation}
where $\delta m^2$ is a  positive constant and $m$ is the quark mass.
This behavior, however, is incorrect, and the correct one is given as follows.

By minimizing the potential  with the vacuum ansatz  that $U(x) = U_0 = {\rm e}^{i\varphi_0}$, we obtain
\begin{equation}
\varphi_0 = \left\{
\begin{array}{cc}
 0 , &  m+\Delta > 0    \\
 \pi,   &   m+\Delta  < 0  \\
\end{array}
\right. ,
\end{equation}
and in terms of the fluctuation around this vacuum that  $U(x) = U_0 {\rm e}^{i\pi (x)/f}$, we have
\begin{eqnarray}
L &=& \frac{1}{2} \left[ \partial_\mu \pi(x) \partial^\mu \pi +m_{\rm PS}^2 \pi(x)^2\right] + O(\pi^4) ,
\quad
m_{\rm PS}^2 \equiv \frac{\vert m +\Delta \vert }{f^2} ,
\end{eqnarray}
which shows that $m_{\rm PS}^2$ is non-symmetric under $m\rightarrow -m$  and becomes zero at $m=-\Delta$.

\subsection{$N_f=2$ case} 
For $N_f=2$, without a loss of generality, the mass term is taken as
\begin{equation}
M = {\rm e}^{i\theta} \left(
\begin{array}{cc}
 m_u & 0 \\
0 &  m_d \\
\end{array}
\right)
\equiv
{\rm e}^{i\theta} 2 B \left( 
\begin{array}{cc}
m_{0u} & 0 \\
0 &  m_{0d} \\
\end{array}
\right) , 
\end{equation}
where $B$ is a positive and mass-independent constant related to the chiral condensate in the massless limit, $m_{0u,0d}$ are bare quark masses, and $\theta$ corresponds to the vacuum angle of QCD. We take $\theta=0$ unless otherwise stated.

\begin{figure}[tbh]
\begin{center}
\scalebox{0.25}{
\includegraphics
{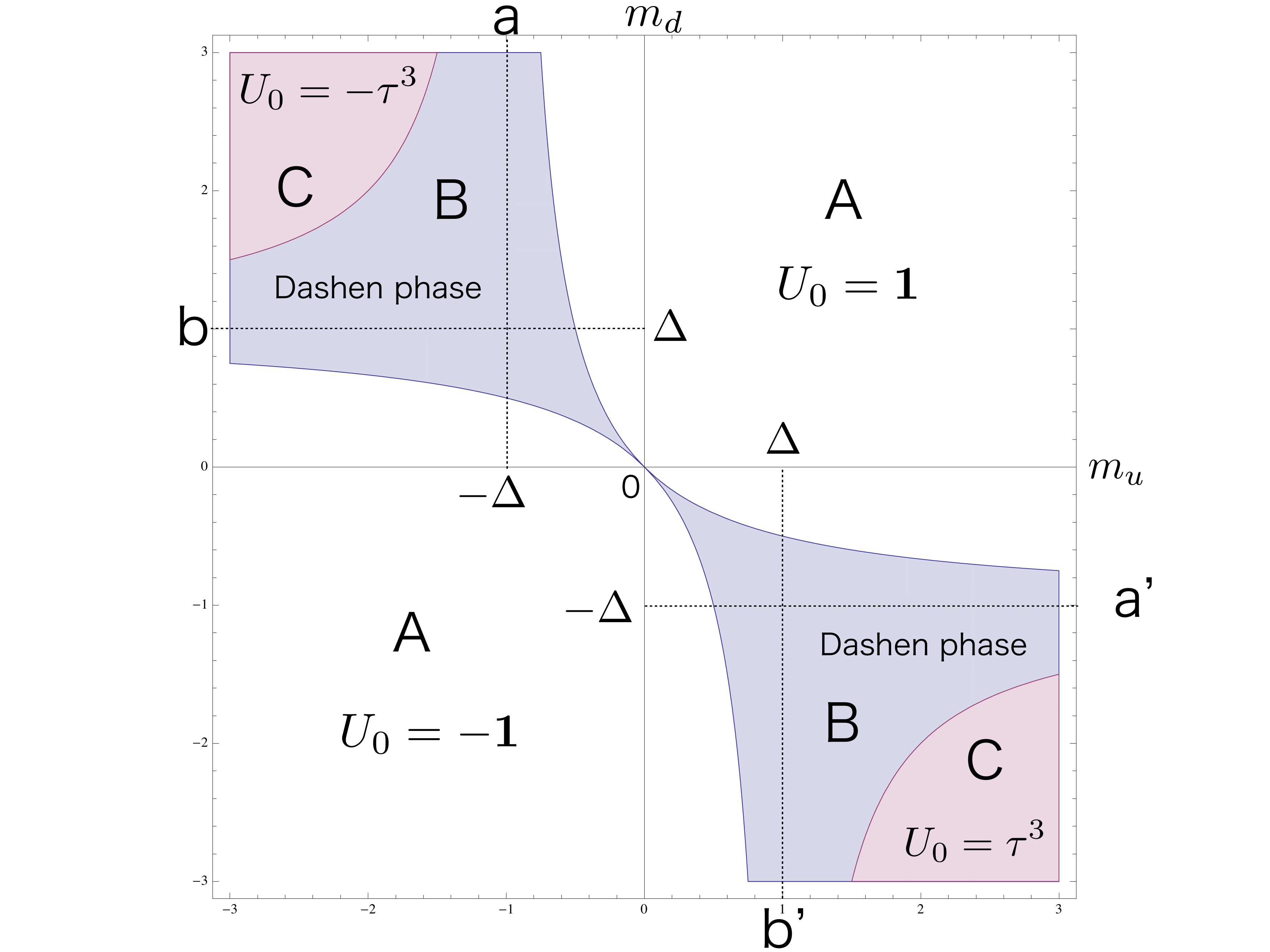}}\hfill
\end{center}
\caption{Phase structure in $m_u$-$m_d$ plain, where  the CP breaking Dashen phase are shaded in blue, while the CP preserving phase with $U_0=\tau^3$ (lower right) or $U_0=-\tau^3$ (upper left) are shaded in red. }
\label{fig:phase}
\end{figure}

Minimizing the potential with the ansatz 
\begin{equation}
U(x) = U_0 ={\rm e}^{i\varphi_0} \left(
\begin{array}{cc}
{\rm e}^{i\varphi_3} & 0 \\
0 &  {\rm e}^{-i\varphi_3} \\
\end{array}
\right),
\end{equation}
we obtain the phase structure in Fig.~\ref{fig:phase}, which is symmetric with respect to $m_+\equiv m_u+m_d=0$ axis an $m_-\equiv m_d-m_u=0$ axis, separately.

In the phase A, $U_0={\bf 1}_{2\times 2}$ (upper right) or $U_0=-{\bf 1}_{2\times 2}$ (lower left), 
while in the phase C, $U_0=\tau^3 $ (lower right) or $U_0=-\tau^3$ (upper left).
Note that the phase $C$ does not exist if we use $ (\log \det\, U)^2$ for the anomaly term, so that an appearance of the phase $C$ could be an artifact of $\det\, U$ term in eq.~(\ref{eq:Lag}).
In the phase B, on the other hand, nontrivial minimum appears as
\begin{eqnarray}
\sin^2(\varphi_3) &=& \frac{ (m_d-m_u)^2\{(m_u+m_d)^2\Delta^2 -
  m_u^2m_d^2\}}{4m_u^3m_d^3} \\ 
 \sin^2(\varphi_0) &=& \frac{
  (m_u+m_d)^2\Delta^2 - m_u^2m_d^2}{4m_u m_d\Delta^2}, 
\end{eqnarray}
which breaks CP symmetry spontaneously, leading to the Dashen phase\cite{Dashen:1971aa},
whose boundaries correspond to the 2nd-order phase transition lines.
Conditions that  $(m_d+m_u) \Delta + m_d m_u = 0$ (a line $\overline{a a^\prime}$) and
$(m_d+m_u) \Delta - m_d m_u = 0$ (a line $\overline{b b^\prime}$) determine
lines between phase A and phase B, 
while those between B and C, on which
$\sin^2\varphi_3=\sin^2\varphi_0 = 1$, are given by  
$(m_d-m_u)\Delta + m_d m_u = 0$ (a line $\overline{ab}$) and $(m_d-m_u)\Delta - m_d m_u =
0$ (a line $\overline{a^\prime b^\prime}$). 

Let consider  PS mesons, described by $U(x) = U_0 {\rm e}^{i \Pi(x)/f}$
with
\begin{equation}
\Pi(x) = \left(
\begin{array}{cc}
\displaystyle\frac{\eta(x)+\pi_0(x)}{\sqrt{2}} & \pi_-(x) \\
\pi_+(x) & \displaystyle\frac{\eta(x)+\pi_0(x)}{\sqrt{2}} \\
\end{array}
\right),
\end{equation}
whose masses are given by
\begin{equation}
m_{\pi^\pm}^2 = \frac{m_+(\vec\varphi)}{2f^2}, \quad 
m_{\tilde\pi^0}^2 = \frac{1}{2f^2}\left[m_+(\vec\varphi)+\delta m - X\right],\quad
m_{\tilde\eta}^2 = \frac{1}{2f^2}\left[m_+(\vec\varphi)+\delta m + X\right],
\end{equation}
where $X=\sqrt{m_(\vec\varphi)^2+(\delta m)^2}$, $\delta m = 2\Delta \cos(\varphi_0) $, and
$
m_\pm(\vec\varphi) = m_d\cos(\varphi_0-\varphi_3) \pm m_u\cos(\varphi_0+\varphi_3)$.
Note that an inequality  $m_{\tilde\pi^0} \le m_{\pi^\pm}$ is always satisfied. 
It is easy to see that $m_{\tilde\pi^0}^2=0$ at all boundaries of the Dashen phase, showing that the phase transition is of second order. 

In this section we have shown that (1) the spontaneous CP violating phase (Dashen phase) by the pion condensation appears in non-degenerate 2-flavor QCD, (2) the massless neutral pion appears at the phase boundaries, and (3) nothing special happen at $m_u=0$ as long as $m_d\not= 0$.

\section{Topological susceptibility and massless up quark}
In ref.~\cite{Creutz:1995wf, Creutz:2003xu,Creutz:2003xc,Creutz:2013xfa},
it is also claimed that $\chi =\infty$ at the phase boundaries while $\chi=0$ at $m_u=0$ for non-zero $m_d$,
where
\begin{equation}
\chi  \equiv \int d^4 x \langle q(x) q(0) \rangle 
\end{equation}
is the topological susceptibility, and $q(x)$ is the topological charge density.
In this section, we confirm this claim using our ChPT analysis.

In our ChPT, the topological charge density is expressed as
\begin{equation}
2N_f q(x) = \Delta \left\{ \det\, U(x) - \det\, U^\dagger(x) \right\},
\end{equation}
while the topological susceptibility is given by
\begin{equation}
2N_f \chi = \frac{\Delta^2}{4} \int d^4x\,  \langle \left\{ \det\, U(x) - \det\, U^\dagger(x) \right\} \left\{ \det\, U(0) - \det\, U^\dagger(0) \right\}\rangle + \frac{\Delta}{2}\langle \det\, U(0) - \det\, U^\dagger(0) \rangle,
\end{equation}
where the second term represents an effect of the contact term in the chiral WTI\cite{Aoki:2014moa}.
For this formula, we obtain
\begin{equation}
2N_f\chi = - \frac{4\Delta^2 m_+(\vec\varphi)}{m_+^2(\vec\varphi) - m_-^2(\vec\varphi)+ 2m_+(\vec\varphi)\delta m} + \Delta .
\label{eq:chi}
\end{equation}
At $m_u=0$, we have $m_+(\vec\varphi)= m_-(\vec\varphi)=m_d$ and $\delta m = 2\Delta$, which lead to
\begin{equation}
2N_f\chi = -\frac{4\Delta^2 m_d}{4m_d\Delta} +\Delta = 0,
\end{equation}
showing that the contact term is required to make $\chi$ vanish at $m_u=0$. On the other hand, at $m_{\pi^0}^2\rightarrow 0$, the denominator of eq.~(\ref{eq:chi}) vanishes, so that $\chi \rightarrow\infty$.
We thus confirm that the claim mentioned in the beginning of this section.

\section{An interesting application: Degenerate 2-flavor QCD at $\theta =\pi$}
As an application of our analysis, let us consider the 2-flavor QCD with $m_u=m_d=m$ but $\theta=\pi$,
which is shown to equivalent to the $m_u= - m_d$ with $\theta = 0$ by the chiral rotation.
Both systems have a SU(2) symmetry generated by $\{\tau^1,\tau^2,\tau^3\}$ for the former or $\{\tau^1\gamma_5,\tau^2\gamma_5,\tau^3\}$ for the latter. In this section, we analyze the former system but a reinterpretation of results for the latter case is easy. 

In this case, the vacuum expectation values are given by
\begin{eqnarray}
\langle \bar\psi i\gamma_5 \psi \rangle  &=&
\left\{
\begin{array}{lc}
0, &  m^2 \ge 4\Delta^2 \\
\pm 2\sqrt{1-\displaystyle\frac{m^2}{4\Delta^2}},  &  m^2 < 4\Delta^2 \\
\end{array}
\right. , \\
\langle \bar\psi \psi \rangle &=&\left\{
\begin{array}{lc}
2, &  2\Delta \le m \\
\displaystyle\frac{m}{\Delta} ,  & -2\Delta < m < 2\Delta \\
-2,  & m \le -2\Delta \\
\end{array}
\right. , 
\end{eqnarray}
showing the spontaneous CP symmetry breaking at $m^2 < 4 \Delta^2$ by the $\eta$ condensation.
Note that if  $ (\log \det\, U)^2$ term were used for the anomaly, the CP symmetry would  always be spontaneously broken:
The phase transition point at $m^2=4\Delta^2$ in the above  would move to $m^2=\infty$.

\begin{figure}[tbh]
\begin{center}
\scalebox{0.23}{\includegraphics{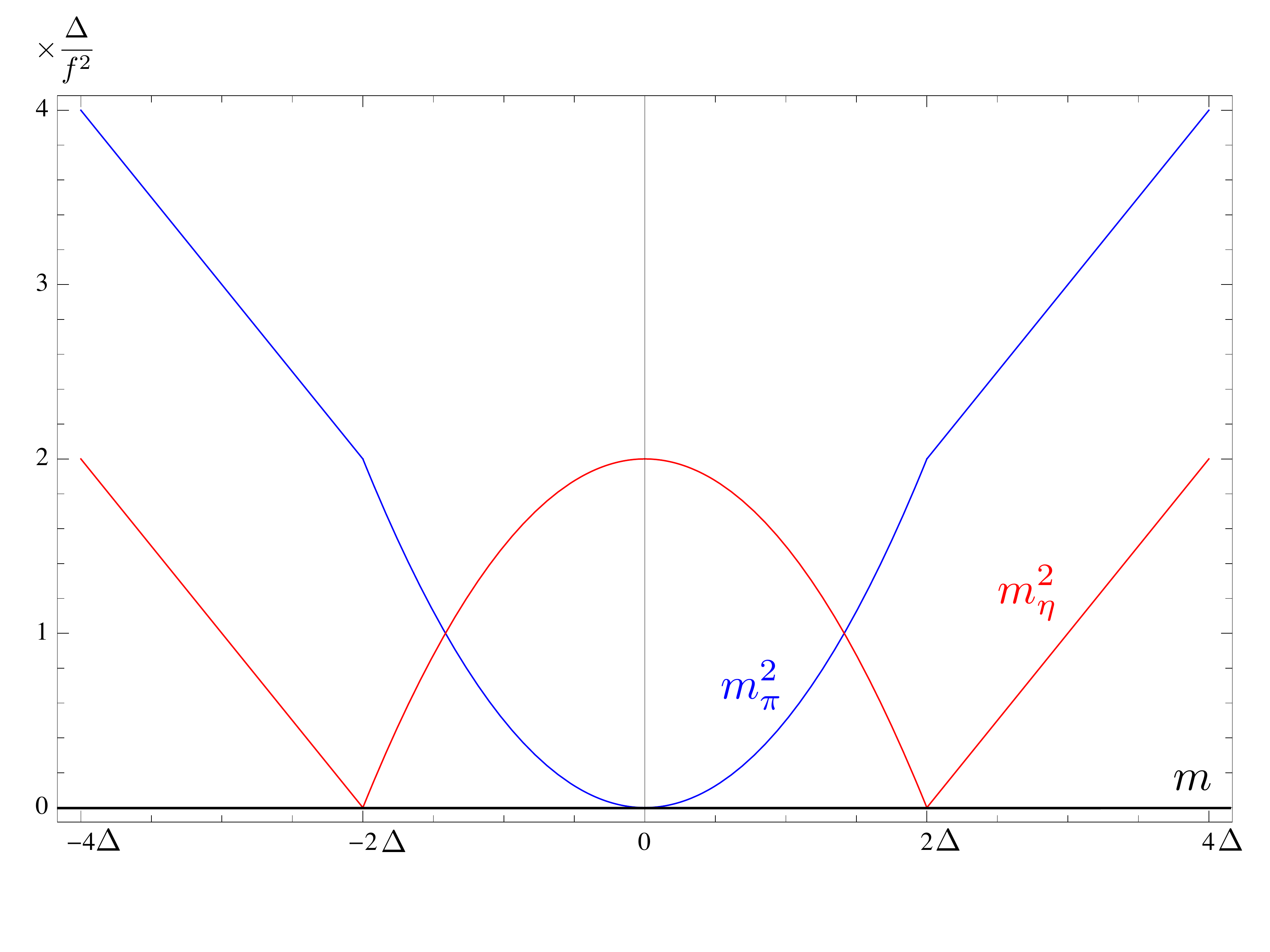}}
\end{center}
\caption{$m_\pi^2$ (blue) and $m_\eta^2$ (red) in unit of $\displaystyle\frac{\Delta}{f^2}$ as a function of $m$.}
\label{fig:PSmesons}
\end{figure}

The PS meson masses are given by
\begin{eqnarray}
m_\pi^2&=& m_{\pi_\pm}^2 = m_{\pi_0}^2 = \left\{
\begin{array}{ll}
\displaystyle\frac{1}{2f^2} 2\vert m\vert, & m^2 \ge 4\Delta^2 \\
\displaystyle\frac{1}{2f^2}\frac{m^2}{\Delta}, & m^2 < 4\Delta^2 \\  
\end{array}
\right. ,\\
m_\eta^2 &=&  \left\{
\begin{array}{ll}
\displaystyle\frac{1}{2f^2} \left[2\vert m\vert-4\Delta \right], & m^2 \ge 4\Delta^2 \\
\displaystyle\frac{1}{2f^2}\frac{4\Delta^2 - m^2}{\Delta}, & m^2 < 4\Delta^2 \\  
\end{array}
\right. ,
\end{eqnarray}
where the $\eta$ meson becomes massless at the phase boundaries at $m^2=4\Delta^2$.
The above mass formulae show that $\eta$ is the massless mode associated with the spontaneous CP violating phase transition while three pions becomes massless NG boson at $m=0$, as given in Fig.~\ref{fig:PSmesons}.
 
 Although the results near the phase transition point around $m^2=4\Delta^2$ may strongly depend on the choice of the anomaly term in our ChPT, properties near $m=0$ can be trusted. Namely, the CP symmetry is spontaneously broken due to the $\eta$ condensation and three pions become massless NG bosons at $m=0$.
It is noted, however, that pion masses behave as $m_\pi^2 = m^2/(2f^2\Delta)$ near $m=0$, in contrast to  the standard PCAC relation that $m_\pi^2 = \vert m\vert/(2f^2)$. 
We now show that this behavior is indeed derived by the WTI, which reads
\begin{eqnarray}
 m \int d^4x\,
{\rm tr}\,\tau^3 ( U^\dagger - U)(x) {\rm tr}\,\tau^3 ( U^\dagger - U)(y)\rangle &=& -2\langle {\rm tr}\, (U+U^\dagger)(y)\rangle .
\end{eqnarray}
From this we obtain
\begin{equation}
m_{\pi_0}^2 = \frac{m}{f^2} \cos\varphi_0 = \frac{m}{f^2}\frac{m}{2\Delta} ,  
\end{equation}
which tells us that one $m$ comes explicitly from the WTI while the other $m$ appears from  $\langle \bar\psi\psi\rangle $. It is interesting but challenging to confirm this behavior by lattice QCD simulations with $\theta = \pi$.

\section{Conclusion}
Using the ChPT including $\eta$ and  the anomaly effect, we have confirmed properties of QCD claimed in Refs.~\cite{Creutz:1995wf, Creutz:2003xu,Creutz:2003xc,Creutz:2005gb,Creutz:2013xfa}:
\begin{enumerate}
\item Nothing is singular at $m_u=0$ if $m_d\not=0$ since no symmetry exists at this point.
\item  The neutral pion becomes massless at  some non-zero values of $m_u$, denoted as $m_u = m_c^\pm$.
\item  The neutral pion condensates ( $\langle \pi^0 \rangle \not= 0$ ) at $m_c^- < m_u < m_c^+$, showing that the CP symmetry is spontaneously broken  ( the Dashen phase ).
Since $m_c^\pm < 0$ for $ m_d > 0$, the Dashen phase can not be investigated by  the staggered quarks with
the rooted trick.
\item The topological susceptibility diverges at $m_\mu = m_c^\pm$.
\item The topological susceptibility vanishes at $m_u=0$, which may give a solution to the strong CP problem.
\end{enumerate}

In addition, we have made new predictions for 2-flavor QCD with $m_u=m_d$ but $\theta = \pi$ as follows.
\begin{enumerate}
\item The spontaneous CP violation occurs as $\langle \eta\rangle \not= 0$.
\item The WTI implies non-standard PCAC relation that $m_\pi^2 \propto m_q^2$. 
\end{enumerate}

\begin{acknowledgments}
S.A thanks Dr. T. Hatsuda  for useful comments.
S.A  is partially supported by Grant-in-Aid
for Scientific Research on Innovative Areas(No.2004:20105001) and
for Scientific Research (B) 25287046 and SPIRE (Strategic Program for Innovative REsearch).
\end{acknowledgments}


\begin{thebibliography}{99}
\bibitem{Nelson:2003tb} 
D.~R.~Nelson, G.~T.~Fleming and G.~W.~Kilcup,
  Phys.\ Rev.\ Lett.\  {\bf 90}, 021601 (2003)
  [hep-lat/0112029].
  
\bibitem{Creutz:1995wf} 
  M.~Creutz,
  Phys.\ Rev.\ D {\bf 52}, 2951 (1995)
  [hep-th/9505112].
  
\bibitem{Creutz:2003xu} 
  M.~Creutz,
  Phys.\ Rev.\ Lett.\  {\bf 92}, 201601 (2004)
  [hep-lat/0312018].

\bibitem{Creutz:2003xc} 
  M.~Creutz,
  Phys.\ Rev.\ Lett.\  {\bf 92}, 162003 (2004)
  [hep-ph/0312225].

\bibitem{Creutz:2005gb}
M. Creutz, 
 {\em PoS}, LAT2005:119 (2006).
  
\bibitem{Creutz:2013xfa} 
  M.~Creutz,
  Annals Phys.\  {\bf 339}, 560 (2013)
  [arXiv:1306.1245 [hep-lat]].  

\bibitem{Dashen:1971aa} 
R.~F.~Dashen,
  Phys.\ Rev.\ D {\bf 3}, 1879 (1971).

\bibitem{Aoki:2014moa} 
  S.~Aoki and M.~Creutz,
  Phys.\ Rev.\ Lett.\  {\bf 112}, 141603 (2014)
  [arXiv:1402.1837 [hep-lat]].

\bibitem{Witten:1980sp} 
  E.~Witten,
  Annals Phys.\  {\bf 128}, 363 (1980).

\bibitem{Rosenzweig:1979ay}
C.~Rosenzweig, J.~Schechter, and C.G. Trahern,
Phys.\ Rev.\ D21 (1980)3388.

\bibitem{Kawarabayashi:1980dp}
Ken Kawarabayashi and Nobuyoshi Ohta.
Nucl.\ Phys.\ B175 (1980) 477.

\bibitem{Arnowitt:1980ne}
Richard~L. Arnowitt and Pran Nath.
Nucl.\ Phys.\ B209 (1982) 234.

\end{thebibliography}
\end{document}